\documentstyle[preprint,eqsecnum,aps]{revtex}
\tightenlines
\newcommand{\be}{\begin{equation}}
\newcommand{\ee}{\end{equation}}
\newcommand{\bea}{\begin{eqnarray}}
\newcommand{\eea}{\end{eqnarray}}
\newcommand{\ba}{\begin{array}}
\newcommand{\ea}{\end{array}}

\newcommand{\th}{\theta}
\newcommand{\Ga}{\Gamma}

\newcommand{\la}{\lambda}
\newcommand{\om}{\omega}

\newcommand{\pa}{\partial}

\newcommand{\no}{\nonumber}

\newcommand{\res}{\mbox{res}}

\begin{document}
\draft
\title{Q-deformed KP hierarchy: Its additional symmetries and\\
infinitesimal B\"acklund transformations}
\author{Ming-Hsien Tu$^1$}

\address{
$^1$ Department of Mathematics, 
National Chung Cheng University, \\
Minhsiung, Chiayi 621, Taiwan}
\date{\today}
\maketitle
\begin{abstract}
We study the additional symmetries associated with the 
$q$-deformed Kadomtsev-Petviashvili ($q$-KP) hierarchy. 
After identifying the resolvent operator as the generator of the additional 
symmetries, the $q$-KP hierarchy can be consistently reduced to the 
so-called $q$-deformed constrained KP ($q$-cKP) hierarchy. 
We then show that the additional symmetries acting on the wave function 
can be viewed as infinitesimal B\"acklund transformations by acting the 
vertex operator on the tau-function of the $q$-KP hierarchy. 
This establishes the Adler-Shiota-van Moerbeke formula for the $q$-KP hierarchy.
\end{abstract}
\newpage


\section{Introduction}

Recently the so-called $q$-deformed Kadomtsev-Petviashvili ($q$-KP) 
hierarchy has being a subject of intensive study in the literature 
\cite{Z,WZZ,JM,F,FR,KLR,HI,Ia,Ib,AHM}. 
The deformation is 
performed to the Lax formulation by introducing a parameter $q$ such that the 
deformed system recovers the ordinary KP hierarchy as $q$ goes to 1. 
Furthermore several integrable structures associated with the ordinary KP 
hierarchy \cite{DJ,Dic} are also maintained in the $q$-KP hierarchy, such as infinite 
conservation laws \cite{WZZ}, bi-Hamiltonian structure \cite{JM}, 
Virasoro and $W$-algebras \cite{JM,FR}, etc.
More recently, the formulation in the spirit of Sato and his school  has been 
carried out for the $q$-KP hierarchy and the concepts of vertex 
operators and tau-functions can be established \cite{HI,Ib,AHM} as well. 

In this letter, we would like to study another interesting property of the
$q$-KP hierarchy, called additional symmetries which have been  connected
to the ordinary KP hierarchy from different points of view such as
conformal algebras \cite{OS} and string equation of matrix models in the 2d quantum
gravity (see, e.g. Refs.\cite{AM,ASM,D}), etc.. 
For the ordinary KP hierarchy, Adler, Shiota and van Moerbeke 
have shown \cite{ASM} that the additional symmetries acting on the wave function
is equivalent to  infinitesimal B\"acklund transformations (BTs) acting on the corresponding
tau-function of the KP hierarchy. The proof was then simplified by Dickey 
using the resolvent operator \cite{D1}. Therefore it is quite interesting to 
see whether we can  carry out this correspondence in the $q$-deformed framework.

Our main results contain two part. 
The first part (Sec. III.) is to construct the generator of the
additional symmetries using the resolvent operator. As a by-product of this discussion,
 we obtain a reduction of the $q$-KP hierarchy, called $q$-deformed constrained KP 
($q$-cKP) hierarchy. The second part (Sec. IV.) is devoted to the proof of the 
Adler-Shiota-van Moerbeke formula for the $q$-KP hierarchy.

Since our approach relies on the use of $q$-deformed pseudodifferential
operators ($q$-PDOs ), let us recall some basic facts about 
$q$-PDOs in the following. 

The $q$-deformed derivative $\pa_q$ is defined by
\be
(\pa_qf(x))=\frac{f(qx)-f(x)}{x(q-1)}
\label{qd}
\ee
which recovers the ordinary differentiation $(\pa_xf(x))$ as $q$ goes to 1.
Let us define the $q$-shift operator $\th$ such that
$\th(f(x))=f(qx)$ and then we have $(\pa_q\th^k(f))=q^k\th^k(\pa_qf),( k\in Z) $.
 Further, from the definition (\ref{qd}), we have  
$(\pa_qfg)=(\pa_qf)g+\th(f)(\pa_qg)$ which implies that
$\pa_qf=(\pa_qf)+\th(f)\pa_q$. Let $\pa_q^{-1}$ denote the formal 
inverse of $\pa_q$ then 
$\pa_q^{-1}f=\sum_{k\geq 0}(-1)^kq^{-k(k+1)/2}\th^{-k-1}(\pa_q^kf)
\pa_q^{-k-1}$.

In general the following $q$-deformed Leibniz rule holds.
\be
\pa_q^nf=\sum_{k\geq 0}{n\choose k}_q\th^{n-k}(\pa_q^kf)\pa_q^{n-k}\qquad
n\in Z
\ee
where the $q$-number and the $q$-binomial are defined by
\bea
(n)_q&=&\frac{q^n-1}{q-1}\no\\
{n\choose k}_q&=&\frac{(n)_q(n-1)_q\cdots(n-k+1)_q}
{(1)_q(2)_q\cdots(k)_q},\qquad {n\choose 0}_q= 1.
\eea
For a $q$-PDO of the form $P=\sum_{i=-\infty}^np_i\pa_q^i$
it is convenient to separate $P$ into the differential part 
$P_+=\sum_{i\geq 0}p_i\pa_q^i$ and the integral part 
$P_-=\sum_{i\leq -1}p_i\pa_q^i$, respectively. 
The residue of $P$ is defined by $\res_{\pa_q}(P)=p_{-1}$ and the conjugate 
operation ``$*$" for $P$ is defined by $P^*=\sum_i(\pa_q^*)^ip_i$
with $\pa_q^*=-\pa_q\th^{-1}$. Then a straightforward calculation shows that 
$(PQ)^*=Q^*P^*$ where $P$ and $Q$ are any two $q$-PDOs.


\section{$q$-deformed KP hierarchy}

The $q$-KP hierarchy is defined by the Lax equation
\be
\pa_nL=[L^n_+, L],\qquad \pa_n\equiv \frac{\pa}{\pa t_n}
\label{laxeq}
\ee
with Lax operator of the form
\be
L=\pa_q+\sum_{i=0}^{\infty}u_i\pa_q^{-i}.\no
\ee
According to the Sato theory, we can express the Lax operator 
as a dressed operator
\be
L=S\pa_q S^{-1}\no
\ee
where $S=1+\sum_{i=1}^{\infty}\om_i\pa_q^{-i}$ is called the Sato operator
and $S^{-1}$ is its formal inverse.
In terms of $S$ the Lax equation (\ref{laxeq}) is equivalent to the Sato equation
\be
\pa_nS=-(L^n)_-S.
\label{satoeq}
\ee
Let us introduce the $q$-deformed exponential function
\be
e_q^x=\exp\left[\sum_{k=1}^{\infty}\frac{(1-q)^k}{k(1-q^k)}x^k\right]\no
\ee
which implies  that $(e_q^x)^{-1}=e_{1/q}^{-x}$,  
$\pa_q e_q^{xz}=ze_q^{xz}$ and  $q\pa_q^* (e_q^{xz})^{-1}=z(e_q^{xz})^{-1}$.

Using $S$ we can define the wave function $\om_q$ and 
adjoint function $\om_q^*$, respectively as follows
\bea
\om_q&=&Se_q^{xz}e^{\xi(t,z)}=
(1+\sum_{i=1}^{\infty}\om_iz^{-i})e_q^{xz}e^{\xi(t,z)}
=\hat{\om}_qe_q^{xz}e^{\xi(t,z)},\no\\
\om^*_q&=&(S^*)^{-1}|_{x/q}(e_q^{xz})^{-1}e^{-\xi(t,z)}
=(1+\sum_{i=1}^{\infty}\om^*_iz^{-i})(e_q^{xz})^{-1}e^{-\xi(t,z)}
=\hat{\om}_q(e_q^{xz})^{-1}e^{-\xi(t,z)}.
\eea
where $\xi(t,z)=\sum_{i=1}^{\infty}t_iz^i$ and $P|_{x/q}=\sum_i p_i(x/q)q^i\pa_q^i$.
It is easy to show that $\om_q$ and $\om_q^*$ satisfy the following linear systems
\bea
L\om_q&=&z\om,\qquad \pa_n\om=(L^n)_+\om\no\\
(L^*|_{x/q})\om_q^*&=&z\om_q^*,\qquad 
\pa_n\om_q^*=-(L^*|_{x/q})^n_+\om_q^*.
\eea

{\bf Lemma 2.1.\/}\cite{Ib} Let $P$ and $Q$ be two $q$-PDOs. Then
\be
\res_z(Pe_q^{xz}Q^*|_{x/q}(e_q^{xz})^{-1})=\res_{\pa_q}(PQ)\no
\ee
where we denote two types of residues as $\res_z(\sum_ia_iz^i)=a_{-1}$ 
and $\res_{\pa_q}(\sum_ia_i\pa_q^i)=a_{-1}$, respectively.

{\bf Lemma 2.2.\/}(see, e.g. Sec. 7.7 of Ref.\cite{Dic}) 
If $f(z)=\sum_{-\infty}^{\infty}a_iz^{-i}$ then
\be
\res_z[\zeta^{-1}(1-z/\zeta)+z^{-1}(1-\zeta/z)]f(z)=f(\zeta)\no
\ee
where $(1-z/\zeta)$ and $(1-\zeta/z)$ should be understood as series in 
$\zeta^{-1}$ and $z^{-1}$, respectively.

The $q$-KP hierarchy admits a reduction defined by $(L^n)_-=0$.
This reduction is compatible with the hierarchy flows because
\be
\pa_k(L^n)_-=[(L^k)_+, L^n]_-=[(L^k)_+, (L^n)_-]_-.\no
\ee
Hence $(L^n)_-$ is identical to zero if its initial value is.
The hierarchy associated with this reduction is called $q$-deformed
$n$-th KdV hierarchy whose Lax operator is thus defined by the
$n$-th order $q$-deformed differential operator $(L^n)_+$.


\section{additional symmetries}

Let us turn to the additional symmetries of the $q$-KP hierarchy.
Our approach follows closely that of Dickey \cite{D1} by considering
the Lax equation (\ref{laxeq}) as a dressing of the commutation relation
$[\pa_k-\pa_q^k, \pa_q]=0$. On the other hand, we find another operator
\be
\Ga_q=\sum_{i=1}^{\infty}\left[it_i+\frac{(1-q)^i}{(1-q^i)}x^i\right]\pa_q^{i-1}\no
\ee
which also commutes with $\pa_k-\pa_q^k$, i.e., $[\pa_k-\pa_q^k, \Ga_q]=0$.
Dressing it gives 
\be
\pa_kM=[(L^k)_+, M],\qquad M\equiv S\Ga_q S^{-1}
\label{meq}
\ee
which are hierarchy flows for the operator $M$. Eq.(\ref{laxeq}) together 
with (\ref{meq}) implies that
\be
\pa_k(M^mL^l)=[(L^k)_+, M^mL^l].
\label{mleq}
\ee
In contrast to the Sato equation (\ref{satoeq}), we define the additional flow 
for each pair $m$, $l$ as follows 
\be
\pa^*_{ml}S=-(M^mL^l)_-S,\qquad \pa^*_{ml}\equiv \frac{\pa}{\pa t^*_{ml}}\no
\ee
Then
\be
\pa^*_{ml}L=-[(M^mL^l)_-, L],\qquad
\pa^*_{ml}M=-[(M^mL^l)_-, M].\no
\ee
Hence $(M^mL^l)_-$ serves as the generator of the additional symmetries
along the trajectory parametrized by $t^*_{ml}$. Moreover it can 
be easily shown that the additional flows $\{\pa^*_{ml}\}$ commute 
with the hierarchy flows, i.e., $[\pa^*_{ml}, \pa_k]L=0$ but
do not commute with each other due to the fact that the operator
$M$ depends on $t_k$ explicitly. 

Let us linearly combine the vector fields $\pa^*_{ml}$ to form a new
vector field $\pa^*_{\mu\la}=\sum_{m=0}^{\infty}\sum_{l=-\infty}^{\infty}
(\mu-\la)^m\la^{-m-l-1}/m!\pa^*_{m,m+l}$ which also belongs to the 
additional flows. Then the generator associated with this new
vector field is given by
\be
Y_q(\mu,\la)=\sum_{m=0}^{\infty}\frac{(\mu-\la)^m}{m!}
\sum_{l=-\infty}^{\infty}\la^{-m-l-1}(M^mL^{m+l})_-.\no
\ee

{\bf Theorem 3.1.\/}(Dickey) The operator $Y_q(\mu, \la)$ can be expressed as
\be
Y_q(\mu, \la)=\om_q(t,\mu)\pa_q^{-1}\th(\om_q^*(t, \la)).\no
\ee
Before proving the theorem, let us prepare the following identities.

{\bf Lemma 3.2.\/} Let $P$ be a $q$-PDO. 
Then we have
\be
(P)_-=\sum_{i=1}^{\infty}\pa_q^{-i}\th(\res_{\pa_q}(\pa_q^{i-1}P)).\no
\ee
In particular,
\be
f\pa_q^{-1}=\sum_{i=1}^{\infty}\pa_q^{-i}\th(\pa_q^{i-1}f).\no
\ee
{\it Proof of theorem 3.1.\/} 
Using Lemma 2.1 and Lemma 3.2 we have
\bea
(M^mL^{m+l})_-
&=&(S\Ga_q^m\pa_q^{m+l}S^{-1})_-\no\\
&=&\sum_{i=1}^{\infty}\pa_q^{-i}\th[\res_{\pa_q}
(\pa_q^{i-1}S\Ga_q^m\pa_q^{m+l}S^{-1})]\no\\
&=&\sum_{i=1}^{\infty}\pa_q^{-i}\th[\res_z
(\pa_q^{i-1}S\Ga_q^m\pa_q^{m+l}e_q^{xz}e^{\xi(t,z)}
(S^{-1})^*|_{x/q}(e_q^{xz})^{-1}e^{-\xi(t,z)})]\no\\
&=&\res_z\left[\sum_{i=1}^{\infty}\pa_q^{-i}\th(
\pa_q^{i-1}z^{m+l}(\pa_z^m\om_q)\om_q^*)\right]
=\res_z[z^{m+l}(\pa_z^m\om_q)\pa_q^{-1}\th(\om_q^*)]\no
\eea
where we have used $\Ga_q\om_q=\pa_z\om_q$ to reach the last 
second equality. Next, using Lemma 2.2, we have
\bea
Y_q(\mu, \la)&=&\res_z\left[\sum_{m=0}^{\infty}\frac{(\mu-\la)^m}{m!}
\sum_{l=-\infty}^{\infty} \la^{-m-l-1}z^{m+l}(\pa_z^m\om_q)\pa_q^{-1}
\th(\om_q^*)  \right]\no\\
&=&\res_z \left[ (\frac{1}{z(1-\la/z)}+\frac{1}{\la(1-z/\la)})e^{(\mu-\la)\pa_q}
\om_q(t,z)\pa_q^{-1}\th(\om_q^*(t,z)) \right]\no\\
&=&((e^{(\mu-\la)\pa_{\la}}\om_q(t, \la))\pa_q^{-1}\th(\om_q^*(t,\la))
=\om_q(t,\mu)\pa_q^{-1}\th(\om_q^*(t, \la)).\quad \Box \no 
\eea 
We further introduce a generator $Y(t)$ which is constructed from 
$Y_q(\mu, \la)$ by integrating the spectral parameters $\mu$ and $\la$ 
with some weightings. Let $\phi(t)=\int\rho(\mu)\om_q(t,\mu)d\mu$ and 
$\psi(t)=\int\chi(\la)\th(\om_q^*(t,\la))d\la$. 
Then we have $Y(t)=\phi(t)\pa_q^{-1}\psi(t)$ with
\be
\pa_k\phi=(L^k)_+\phi,\qquad \pa_k\psi=-(L^k)^*_+\psi.
\label{eigeneq}
\ee
Introducing new symmetries to a system always induces constraints 
to the system. In our present case, the additional symmetries generated by the
operator $(M^mL^l)_-$ provides another reduction $(L^n)_-=(M^mL^l)_-$
 to the $q$-KP hierarchy. 
From (\ref{laxeq}) and (\ref{mleq}) it is easy to show that 
$\pa_k((L^n)_--(M^mL^l)_-)=[(L^k)_+, ((L^n)_--(M^mL^l)_-)]_-$ 
and hence this reduction is compatible with the hierarchy flows as well.
The situation still holds for the generator
$Y(t)$ since it is a special linear combinations of the operators $(M^mL^l)_-$.
Therefore we obtain a reduction of the $q$-KP hierarchy defined by the 
Lax operator
\be
K=L^n=(L^n)_++\phi(t)\pa_q^{-1}\psi(t)\no
\ee  
which satisfies the Lax equation
\be
\pa_kK=[(K^{k/n})_+, K].
\label{claxeq}
\ee
It is easy to show that (\ref{claxeq}) is compatible with Eq.(\ref{eigeneq}).
We call this hierarchy the $q$-deformed constrained KP 
($q$-cKP) hierarchy which has rich integrable structures as well as
$q$-KP hierarchy\cite{T}.

\section{infinitesimal B\"acklund transformations}

Through the bilinear identity it has been shown \cite{Ib,AHM} that the (adjoint) wave 
functions of the $q$-KP hierarchy can be expressed in terms of a single function 
$\tau_q(x;t)$ called tau-function such that
\be
\hat{\om}_q=\frac{\tau_q(x;t-[z^{-1}])}{\tau_q(x;t)}
\label{wave},\qquad
\hat{\om}_q^*=\frac{\tau_q(x;t+[z^{-1}])}{\tau_q(x;t)}
\label{adjwave}
\ee
where $[z^{-1}]=(1/z, 1/2z^2, 1/z^3, \cdots)$. A peculiar property
of $\tau_q(x;t)$ is that it can be obtained from the tau-function $\tau(t)$ of 
the ordinary KP hierarchy by shifting the time variables $t$ to $t+[x]_q$,
i.e., $\tau_q(x;t)=\tau(t+[x]_q)$ 
where $[x]_q=(x, (1-q)^2x^2/2(1-q^2), (1-q)^3x^3/3(1-q^3), \cdots$).
For convenience, from now on, the tau-function $\tau_q(x;t)$ will be simply written by
$\tau_q(t)$ without any confusion. 

Now let us define the vertex operators
\bea
X_q(t,z)&=&e_q^{xz}\exp\left(\sum_{i=1}^{\infty}t_iz^i\right)
\exp\left(-\sum_{i=1}^{\infty}\frac{z^{-i}}{i}\pa_i\right)\no\\
X^*_q(t,z)&=&(e_q^{xz})^{-1}\exp\left(-\sum_{i=1}^{\infty}t_iz^i\right)
\exp\left(\sum_{i=1}^{\infty}\frac{z^{-i}}{i}\pa_i\right)
\eea
then 
\be
\om_q=\frac{X_q(t,z)\tau_q(t)}{\tau_q(t)},\qquad 
\om_q^*=\frac{X^*_q(t,z)\tau_q(t)}{\tau_q(t)}
\ee
We further define a new vertex operator $X_q(\mu,\la)$ which is
constructed as follows:
\be
X_q(\mu,\la)=e_q^{x\mu}(e_q^{x\la})^{-1}
\exp\left[\sum_{i=1}^{\infty}t_i(\mu^i-\la^i)\right]
\exp\left[-\sum_{i=1}^{\infty}\frac{(\mu^{-i}-\la^{-i})}{i}\pa_i\right].\no
\ee
According to the approach of Sato, it can be proved \cite{AHM} that
the action of $X(\mu, \la)$ on $\tau_q$ corresponds to an infinitesimal
BTs on the space of tau-functions. In other\
words, if $\tau_q$ is a solution of the $q$-KP hierarchy then 
$\tau_q+\epsilon X_q(\mu,\la)\tau_q$ a solution too. 
To connect such infinitesimal BTs to the additional symmetries described above,
we prepare the following two lemmas concerning the $q$-analogue of the
(differential) Fay identity.

{\bf Lemma 4.1.\/} The tau-function $\tau_q$ satisfies the following 
$q$-deformed Fay identity
\be
\sum_{(s_1,s_2,s_3)}(s_0-s_1)(s_2-s_3)\tau_q(t+[s_0]+[s_1])\tau_q(t+[s_2]+[s_3])=0
\label{fay}
\ee
where $(s_1,s_2,s_3)$ denotes cyclic permutations.

The above identity is obtained from the Fay identity for tau-functions
of the ordinary KP hierarchy by shifting the variables $t$ to $t+[x]_q$. 

{\bf Lemma 4.2.\/}  The tau-function $\tau_q$ satisfies the following 
$q$-deformed differential Fay identity
\bea
& &\pa_q\tau_q(t-[s_3]+[s_1])\th(\tau_q(t))-
\th(\tau_q(t-[s_3]+[s_1]))(\pa_q\tau_q(t))\no\\
& &=(s_3^{-1}-s_1^{-1})\th(\tau_q(t+[s_1]))\tau_q(t-[s_3])-
s_3^{-1}\th(\tau_q(t-[s_3]+[s_1]))\tau_q(t)\no\\
& &+s_1^{-1}\tau_q(t-[s_3]+[s_1])\th(\tau_q(t))
\label{dfay}
\eea
{\it Proof of lemma 4.2.\/} From the Fay identity (\ref{fay}),
setting $s_0=0$ and  divided by $s_1s_2s_3$ we have
\be
\sum_{(s_1,s_2,s_3)}(s_3^{-1}-s_2^{-1})\tau_q(t+[s_1])\tau_q(t+[s_2]+[s_3])=0.
\ee
Letting $s_2=(1-q)x$, shifted $t \to t-[s_2]-[s_3]$ and
using the relation $[qx]_q=[x]_q+[(1-q)x]$ then (\ref{dfay}) follows. $\Box$

{\bf Theorem 4.3.\/} (Adler-Shiota-van Moerbeke and Dickey)
The following formula
\be
X_q(\mu, \la)\om_q(t,z)=(\la-\mu)Y_q(\mu, \la)\om_q(t,z)\no
\ee
holds, where it should be understood that the vertex operator $X_q(\mu, \la)$ 
acting on $\om_q$ is generated by its action on $\tau_q$ through the expression
(\ref{wave}). 

{\it Proof of theorem 4.3.\/} From (\ref{wave}) we have
\bea
& & X_q(\mu,\la)\frac{\tau_q(t-[z^{-1}])}{\tau_q(t)}=
\frac{\tau_q(t)e^{-\sum_{i=1}^{\infty}z^{-i}/i\pa_i}X_q(\mu,\la)\tau_q(t)-
\tau_q(t-[z^{-1}])X_q(\mu,\la)\tau_q(t)}{\tau_q^2(t)}\no\\
& &=e_q^{x\mu}(e_q^{x\la})^{-1}e^{\xi(t,\mu-\la)}
\left[\tau_q(t)\tau_q(t-[z^{-1}]-[\mu^{-1}]+[\la^{-1}])(z-\mu)(z-\la)^{-1}\right.\no\\
& &\left.-\tau_q(t-[z^{-1}])\tau_q(t-[\mu^{-1}]+[\la^{-1}]) \right]/\tau_q^2(t)\no\\
& &=e_q^{x\mu}(e_q^{x\la})^{-1}e^{\xi(t,\mu-\la)}(z-\la)^{-1}
\left[(z-\mu)\tau_q(t'+[z^{-1}]+[\mu^{-1}])\tau_q(t'+[\la^{-1}]) \right.\no\\
& &\left. -(z-\la)\tau_q(t'+[\mu^{-1}])\tau_q(t'+[z^{-1}]+[\la^{-1}])\right]/\tau_q^2(t)\no
\eea
here $t'\equiv t-[z^{-1}]-[\mu^{-1}]$. 
Using the Fay identity (\ref{fay}) with  $s_0=0, s_1=\la^{-1}, 
s_2=\mu^{-1}, s_3=z^{-1}$ then the terms in the bracket can be simplified to
$(\la-\mu)\tau_q(t'+[z^{-1}])\tau_q(t'+[\mu^{-1}]+[\la^{-1}])$.

Hence the following connection 
\bea
X_q(\mu,\la)\om_q(t,z)&=&
e_q^{xz}e_q^{x\mu}(e_q^{x\la})^{-1}e^{\xi(t,z+\mu-\la)}\frac{(\la-\mu)}{(z-\la)}
\frac{\tau_q(t-[\mu^{-1}])\tau_q(t-[z^{-1}]+[\la^{-1}])}{\tau_q^2(t)}\no\\
&=&(\la-\mu)\om_q(t,\mu)\pa_q^{-1}\th(\om^*_q(t,\la))\om_q(t,z)\no
\eea
should be established for the proof. 
Multiplying both  sides by $(\la-\mu)^{-1}\pa_q(\om_q(t,\mu))^{-1}$ and 
using the expressions (\ref{wave})  we obtain
\bea
\pa_q\frac{\tau_q(t-[z^{-1}]+[\la^{-1}])}{\tau_q(t)}&=&
(z-\la)\th\left(\frac{ \tau_q(t+[\la^{-1}])}{\tau_q(t)}\right)
\frac{\tau_q(t-[z^{-1}])}{\tau_q(t)}-
z\th\left(\frac{\tau_q(t-[z^{-1}]+[\la^{-1}])}{\tau_q(t)}\right)\no\\
& &+\la\frac{\tau_q(t-[z^{-1}]+[\la^{-1}])}{\tau_q(t)}
\eea
which is nothing but the $q$-deformed differential Fay identity (\ref{dfay})
with $s_1=\la^{-1}$ and $s_3=z^{-1}$.
This completes the proof of the theorem. $\Box$


{\bf Acknowledgment\/}
This work is supported by the National Science Council
of the Republic of China under Grant 
Numbers NSC 88-2811-M-194-0003

\end{document}